\documentclass[apj]{emulateapj}
\usepackage{apjfonts}
\usepackage{graphicx}

\renewcommand{\vec}[1]{\mbox{\boldmath $\displaystyle #1$}}
\newcommand{\grad}{\vec{\nabla}}

\newcommand{\vdot}{\vec{\cdot}}

\newcommand{\divr}{\grad\vdot\,}

\newcommand{\Ye}{Y_{\mathrm{e}}}

\usepackage{color}

\begin{document}
 
\title{Millihertz Quasi-Periodic Oscillations from Marginally Stable
Nuclear Burning on an Accreting Neutron Star}

\author{
Alexander Heger\altaffilmark{1,2}, 
Andrew Cumming\altaffilmark{3}, \& 
Stanford E.~Woosley\altaffilmark{2}}
\altaffiltext{1}{Theoretical Astrophysics Group, T-6, MS B227, 
Los Alamos National Laboratory, Los Alamos, NM 87545; aheger@lanl.gov} 
\altaffiltext{2}{Department of Astronomy and Astrophysics, University
of California, Santa Cruz, CA 95064; alex@ucolick.org, woosley@ucolick.org}
\altaffiltext{3}{Physics Department, McGill University, 3600 rue
University, Montreal, QC, H3A 2T8, Canada; cumming@physics.mcgill.ca}

\begin{abstract}
We investigate marginally stable nuclear burning on the surface of
accreting neutron stars as an explanation for the mHz quasi-periodic
oscillations (QPOs) observed from three low mass X-ray binaries. At
the boundary between unstable and stable burning, the temperature
dependence of the nuclear heating rate and cooling rate almost
cancel. The result is an oscillatory mode of burning, with an
oscillation period close to the geometric mean of the thermal and
accretion timescales for the burning layer. We describe a simple
one-zone model which illustrates this basic physics, and then present
detailed multizone hydrodynamical calculations of nuclear burning
close to the stability boundary using the \textsc{Kepler} code. Our
models naturally explain the characteristic 2 minute period of the mHz
QPOs, and why they are seen only in a very narrow range of X-ray
luminosities. The oscillation period is sensitive to the accreted
hydrogen fraction and the surface gravity, suggesting a new way to
probe these parameters. A major puzzle is that the accretion rate at
which the oscillations appear in the theoretical models is an order of
magnitude larger than the rate implied by the X-ray luminosity when
the mHz QPOs are seen. We discuss the implications for our general
understanding of nuclear burning on accreting neutron stars. One
possibility is that the accreted material covers only part of the
neutron star surface at luminosities $L_X\gtrsim 10^{37}\ {\rm erg\
s^{-1}}$.
\end{abstract}


\section{Introduction}

Low mass X-ray binaries, in which a neutron star or black hole
accretes from a low mass companion, exhibit a range of periodic and
quasi-periodic phenomena, ranging in frequency from very low frequency
(mHz) noise to kHz quasi-periodic oscillations (QPOs) (see van der
Klis 2004 for a recent review). This variability has mostly been
associated with orbiting material in the accretion flow close to the
compact object. In the case of a neutron star accretor, an important
question is whether any of these phenomena originate from or are
associated with the neutron star surface. This is important for
identifying the compact object as a neutron star or a black hole and
offers a probe of the neutron star surface layers.

Unstable nuclear burning on neutron star surfaces has been studied for
many years, and is observed as Type I X-ray bursts. The accreted
hydrogen (H) and helium (He) fuel accumulates on the surface of the
star and undergoes a thin shell instability, giving rise to a
$10$--$100$ second burst of X-rays with typical energy $10^{39}$ ergs
(for reviews, see Lewin, van Paradijs, \& Taam 1993, 1995; Strohmayer
\& Bildsten 2003).  Not all sources show Type I X-ray bursts, however,
and in many sources the bursts are not frequent enough to burn all of
the accreted fuel (van Paradijs, Penninx, \& Lewin 1988; in 't Zand et
al.~2003). Bildsten (1993, 1995) suggested that a different mode of
nuclear burning, involving slowly propagating fires over the neutron
star surface, operates at high accretion rates, and manifests itself
in the power spectrum of the source as very low frequency noise
(VLFN). He found an anti-correlation between bursting and VLFN,
supporting this picture.

Revnivtsev et al.~(2001) discovered a new class of mHz QPOs in three
Atoll sources, 4U~1608-52, 4U~1636-53, and Aql X-1, which they
proposed were from a special mode of nuclear burning on the neutron
star surface rather than from the accretion flow. These mHz QPOs have
frequencies in the range $7$--$9\ {\rm mHz}$ (timescales of
$1.9$--$2.4$ minutes). The associated flux variations are at the few
per cent level, and are strongest at low photon energies ($\lesssim 5$
keV). This is in contrast to all other observed QPOs, whose amplitude
generally increases with photon energy. Revnivtsev et al.~(2001)
showed that the centroid frequency of the mHz QPO was stable on year
timescales in a given source, and the same to within tens of per cent
in all three sources in which it was detected. In addition, the
presence of the mHz QPOs is affected by Type I X-ray bursts: in
4U~1608-52, the mHz QPOs disappeared immediately following a Type I
X-ray burst. In 4U~1608-52, a transient source whose luminosity is
observed to change by orders of magnitude, the mHz QPO was only
present within a narrow range of luminosity, $L_X\approx
0.5$--$1.5\times 10^{37}\ {\rm erg\ s^{-1}}$.

The association of the mHz QPOs with a surface phenomenon was
strengthened by the results of Yu \& van der Klis (2002), who showed
that the kHz QPO frequency is anti-correlated with the luminosity
variations during the mHz oscillation. This is opposite to the long
term trend, which is that the kHz QPO frequency varies proportional to
the X-ray luminosity, consistent with the inner edge of the accretion
disk being pushed inwards at higher accretion rates. The
anti-correlation observed by Yu \& van der Klis (2002) during the mHz
QPO cycle suggests that the inner edge of the disk moves outwards
slightly as the luminosity increases during the cycle, perhaps
consistent with enhanced radiation drag as the gas orbiting close to
the neutron star is radiated by emission from the neutron star
surface.

The fact that the properties of the mHz QPOs suggest that they are
associated with the neutron star surface led Revnivtsev et al.~(2001)
to propose that a special mode of nuclear burning operates at
luminosities $L_X\approx 0.5$--$1.5\times 10^{37}\ {\rm erg\ s^{-1}}$.
The nature of the burning and the physics underlying the
characteristic timescale of $\approx 2$ minutes, however, were
unexplained. Bildsten (1993) gave the characteristic timescales of the
burning layer that might be associated with sub-hertz phenomena. The
thermal timescale of the layer is $t_{\rm therm}\approx 10$ s, the
time to accrete the fuel is $t_{\rm accr}\approx 1000$ s at the
Eddington accretion rate, and the time for a nuclear burning ``fire''
to propagate around the surface is estimated to be $\sim 10^4$
s. Bildsten (1993) proposed that if several fires are propagating around
the star at a given time, the time-signature would be broad band noise
with frequencies in the mHz range.  None of the timescales he
identified match the $\approx 2$ minute mHz QPO period, however.

The luminosity at which mHz QPOs are observed ($L_X\approx 10^{37}\
{\rm erg\ s^{-1}}$) is significant because it is similar to the
luminosity at which a transition in burning behavior occurs, from
frequent Type I X-ray bursting at low accretion rates to the
disappearance of Type I X-ray bursts at high accretion rates. This
transition is common to many X-ray bursters (e.g., Cornelisse et
al.~2003), and is expected theoretically because at high accretion
rates the fuel burns at a higher temperature, reducing the
temperature-sensitivity of helium burning and quenching the thin shell
instability.  An outstanding puzzle is that theory predicts
a transition accretion rate close to the Eddington rate (Bildsten
1998), which corresponds to luminosities $L_X\sim 10^{38}\ {\rm erg\
s^{-1}}$, much larger than observed. Paczynski (1983) pointed out that
near the transition from instability to stability, oscillations are
expected because the eigenvalues of the system are complex. Narayan \&
Heyl (2003) extended Paczynski's one-zone analysis, calculating linear
eigenmodes of truncated steady-state burning models. They too found
complex eigenvalues near the stability boundary, and suggested that
this might explain the mHz QPOs observed by Revnivtsev et al.~(2001).
The oscillation frequencies, however, were an order of magnitude too
small.

In this paper, we show that the mHz QPO frequencies are, in fact,
naturally explained as being due to marginally stable nuclear burning
on the neutron star surface. At the boundary between unstable and
stable burning, the temperature dependence of the nuclear heating rate
and cooling rate almost cancel. The result is an oscillatory mode of
burning, with an oscillation period close to the geometric mean of the
thermal and accretion timescales for the burning layer, $(t_{\rm
therm}t_{\rm accr})^{1/2}\approx 100\ {\rm s}$. In \S 2, we describe a
simple one-zone model which illustrates this basic physics, and then
present detailed hydrodynamical calculations of nuclear burning close
to the stability boundary in \S 3. We discuss the implications of our
results in \S 4. In particular, if the mHz QPOs are due to marginally
stable nuclear burning, the local accretion rate onto the star must be
close to the Eddington rate, even though the global accretion rate
inferred from the X-ray luminosity is ten times lower.

\section{A one-zone model}

In this section, we discuss a simplified model of the burning layers
which illustrates the basic physics underlying the oscillations
observed in our multizone numerical simulations. Following Paczynski
(1983), we consider a one-zone model of the burning layer. The
temperature, $T$, and thickness of the fuel layer, $y$ (which we
measure as a column depth, in units of mass per unit area), obey
\begin{eqnarray}\label{eq:heateqn}
c_P{dT\over dt}&=&\epsilon -{F\over y}
\\\label{eq:conteqn}
{dy\over dt}&=&\dot m - {\epsilon\over E_\star}y.
\end{eqnarray}
(equivalent to eqns.~[8] of Paczynski 1983). Equation
(\ref{eq:heateqn}) describes the heat balance, including heating of
the layer by nuclear reactions $\epsilon$, and radiative cooling
$-\divr\vec{F}/\rho=dF/dy\approx F/y$. The heat capacity at constant
pressure is $c_P$, and $F$ is the outwards heat flux. Equation
(\ref{eq:conteqn}) tracks the burning depth, allowing for accretion of
new fuel at a rate given by the local accretion rate $\dot m$, as well
as burning of fuel on a timescale $E_\star/\epsilon$, where $E_\star$
is the energy per gram released in the burning. Note that the pressure
at the base of the layer is $P=gy$ from hydrostatic balance, where $g$
is the local gravity. We first study equations (\ref{eq:heateqn}) and
(\ref{eq:conteqn}) analytically (\S 2.1), and then show some numerical
integrations (\S 2.2).

\subsection{Analytic  estimates}

Equations (\ref{eq:heateqn}) and (\ref{eq:conteqn}) constitute a
nonlinear oscillator. To understand its behavior, we consider linear
perturbations to the steady state solution, which has
\begin{equation}\label{eq:steadystate}
\epsilon y = F = \dot mE_\star.
\end{equation}
The nuclear energy generation is generally a strong function of
temperature, and we will assume $\epsilon\propto T^\alpha$, where
$\alpha\equiv d\ln\epsilon/d\ln T$. The heat flux is approximately
$F\approx acT^4/3\kappa y$ (e.g., Bildsten 1998). To simplify the
algebra in this section, we assume that $\epsilon$ depends only on
temperature, and that the opacity $\kappa$ is a constant. We write the
deviations from steady state as $\delta y$ and $\delta T$, giving
\begin{eqnarray}
c_P{\partial\delta T\over \partial t}&=&\alpha\epsilon{\delta T\over
T}-{4F\over y}{\delta T\over T}+{2F\over y}{\delta y\over y}\\
{\partial \delta y\over\partial t}&=&-{\epsilon\over
E_\star}\left(\delta y+{\alpha y\over T}\delta T\right).
\end{eqnarray}
Defining a thermal timescale for the layer $t_{\rm
therm}=c_pT/\epsilon=c_pTy/F$ and an accretion timescale $t_{\rm
accr}=y/\dot m$, and using the steady-state relations of equation
(\ref{eq:steadystate}) gives
\begin{eqnarray}\label{eq:pert1}
{\partial\over\partial t}\left({\delta T\over
T}\right)&=&\left({\alpha-4\over t_{\rm therm}}\right){\delta T\over
T}+\left({2\over t_{\rm therm}}\right){\delta y\over y}\\
\label{eq:pert2}
{\partial\over\partial t}\left({\delta y\over
y}\right)&=&-\left({1\over t_{\rm accr}}\right){\delta y\over
y}-\left({\alpha\over t_{\rm accr}}\right){\delta T\over T}.
\end{eqnarray}
It is useful to write $\delta T=f(t)\exp(-t/t_{\rm accr})$ and $\delta
y=g(t)\exp(-t/t_{\rm accr})$. This simplifies equations
(\ref{eq:pert1}) and (\ref{eq:pert2}), allowing us to combine them
into a single differential equation for $f$,
\begin{equation}\label{eq:osc}
{\partial^2f\over\partial t^2}+\left({4-\alpha\over t_{\rm
therm}}-{1\over t_{\rm accr}}\right){\partial f\over \partial
t}+{2\alpha\over t_{\rm accr}t_{\rm therm}}f=0.
\end{equation}
Equation (\ref{eq:osc}) is the equation for a damped simple harmonic
oscillator. The solution is $\delta T\propto\exp(\lambda t)$, with
\begin{equation}
\lambda={1\over 2}\left({\alpha-4\over t_{\rm therm}}-{1\over t_{\rm
accr}}\right) \pm \left[{1\over 4}\left({\alpha-4\over t_{\rm
therm}}+{1\over t_{\rm accr}}\right)^2-{2\alpha\over t_{\rm
therm}t_{\rm accr}}\right]^{1/2}.
\end{equation}
Note that the ``damping'' term in equation (\ref{eq:osc}) can be
positive or negative depending on the relative temperature
sensitivities of the heating and cooling.

\begin{figure}
\plotone{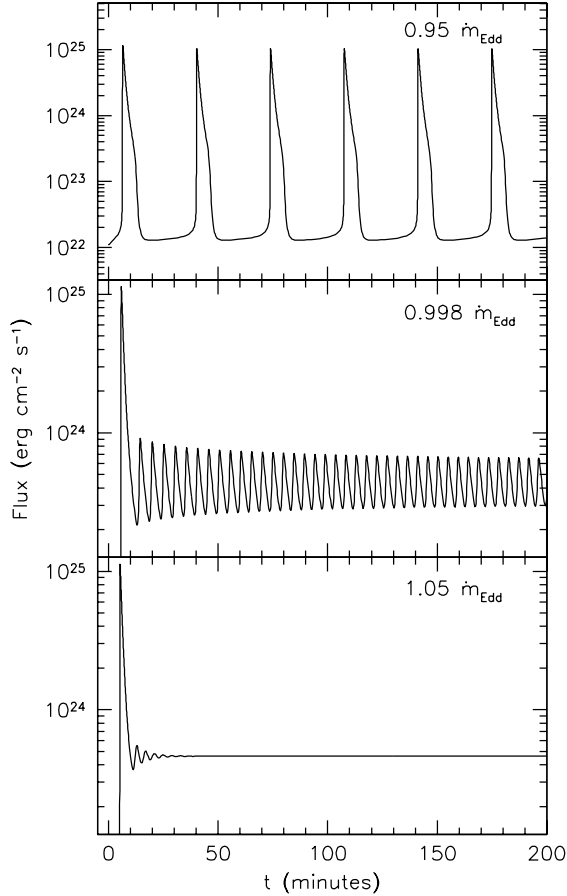}
\caption{Lightcurves for the one-zone model at three different
accretion rates. The system evolves from the arbitrary initial values
$y=2\times 10^8\ {\rm g\ cm^{-2}}$ and $T=2\times 10^8\ {\rm K}$. At
$\dot m=0.95\ \dot m_{\rm Edd}$, bursts occur with a recurrence time
of 34 minutes. At $\dot m=0.998\ \dot m_{\rm Edd}$, oscillations are
seen with a period of 4.7 minutes. At $\dot m=1.05\ \dot m_{\rm Edd}$,
after a few transient oscillations, the burning evolves to a steady
state. The steady state flux is $\dot mE_{\rm nuc}$, where $E_{\rm
nuc}\approx 5$ MeV per nucleon.\label{Fig:lc1}}
\end{figure}

\begin{figure}
\plotone{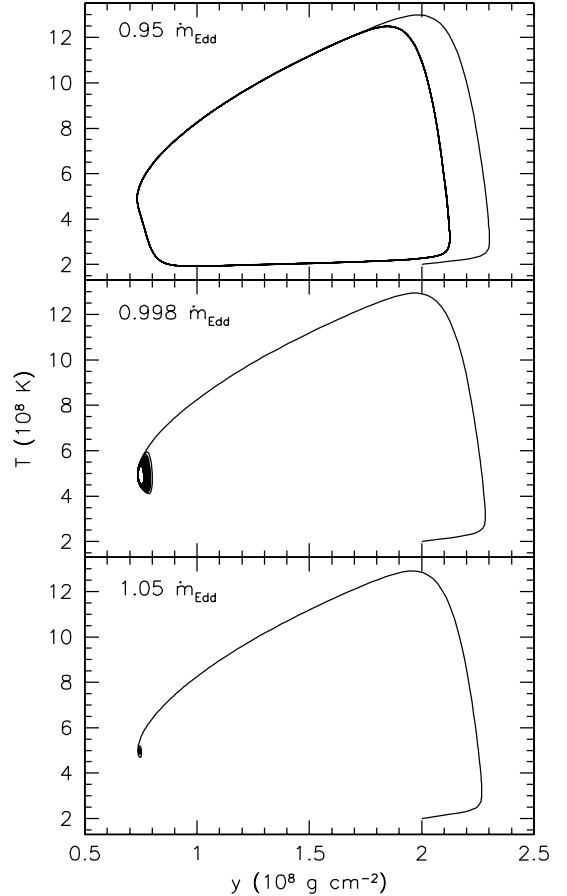}
\caption{The trajectories in the temperature-column depth plane for
the models shown in Figure 1. In each case, the model starts at the
arbitrary point $y=2\times 10^8\ {\rm g\ cm^{-2}}$ and $T=2\times
10^8\ {\rm K}$.\label{Fig:tra1}}
\end{figure}

The key to understanding the behavior of the burning is to note that
the timescales $t_{\rm therm}$ and $t_{\rm accr}$ are very
different. Steady burning at accretion rates near Eddington leads to
ignition at a column depth $y\approx 10^8\ {\rm g\ cm^{-2}}$ and
temperatures $T\approx 5\times 10^8\ {\rm K}$ (Schatz et
al.~1999). The accretion timescale is therefore $y/\dot m\approx 1000\
{\rm s}$ for accretion at $\dot m\approx 10^5\ {\rm g\ cm^{-2}\
s^{-1}}$ (roughly the Eddington rate). For stable burning $F=\dot m
E_\star$, giving $t_{\rm therm}=c_PTy/F=(c_PT/E_\star)(y/\dot
m)\approx 0.01\ t_{\rm accr}\approx 10\ {\rm s}$, since for an ideal
gas, $c_PT\approx (5/2)(k_BT/m_p)=7.2\ {\rm keV}\ T_8$ per nucleon,
whereas $E_\star\approx 5\ {\rm MeV}$ per nucleon (Schatz et
al.~1999)\footnote{In fact, the gas is partially degenerate at the
burning location, with $k_BT\approx E_F$, and so there will be a small
correction factor to the ideal gas expression for $c_P$.}.

Because $t_{\rm therm}\ll t_{\rm accr}$ the damping term is usually
much larger than the oscillatory term, and the system is either
positively or negatively overdamped depending on the sign of
$(\alpha-4)$. When $\alpha>4$, the damping coefficient is negative,
leading to strong exponential growth: the steady state solution is
linearly unstable. When $\alpha <4$, the damping coefficient is
positive, leading to strong damping: the steady-state solution is
linearly stable.  Close to marginal stability when
$\alpha\approx 4$, however, the effective thermal timescale $t_{\rm
therm}/\left|\alpha-4\right|$ becomes much longer than the accretion
time, and the inequality is reversed, giving weakly damped or excited
oscillations with oscillation frequency $\omega\approx (2\alpha/t_{\rm
accr}t_{\rm therm})^{1/2}$. The condition to observe oscillations is
that the oscillation frequency should be larger than the damping rate,
\begin{equation}
\left({2\alpha\over t_{\rm accr}t_{\rm therm}}\right)^{1/2}>{1\over 2}\left({\alpha-4\over t_{\rm therm}}+{1\over t_{\rm accr}}\right).
\end{equation}
Note that this condition can be satisfied when $\lambda$ has either
positive or negative real part, so that the oscillations may be
excited or damped.

Neglecting the slight
modification from the damping term, the oscillation period is $P_{\rm
osc}=2\pi/\omega$, or
\begin{eqnarray}\label{eq:posc}
P_{\rm osc}&\approx &{2\pi\over (2\alpha)^{1/2}}\left({c_PT\over
E_\star}\right)^{1/2}\left({y\over \dot m}\right)\nonumber\\ &=&3.1\
{\rm mins}\ \left({T_8\over 5}\right)^{1/2}y_8\left({\dot m\over \dot
m_{\rm Edd}}\right)^{-1},
\end{eqnarray}
where we set $\alpha=4$ at marginal stability, and take $E_\star=5$
MeV per nucleon. This estimate is remarkably close to the observed mHz
QPO period of $2$ minutes.

The physics of the oscillations can be understood by considering
equations (\ref{eq:pert1}) and (\ref{eq:pert2}). For most accretion
rates, the effective thermal time is much smaller than the accretion
timescale, or $t_{\rm therm}/\left|4-\alpha\right|\ll t_{\rm
accr}$. The perturbations are then effectively at constant pressure,
or $\delta y=0$, as commonly assumed (see, e.g., Fujimoto, Hanawa, \&
Miyaji 1981; Bildsten 1998), and equation (\ref{eq:pert1}) leads
directly to exponential growth or decay depending on the relative
temperature sensitivities of the heating and cooling rates.  Close to
marginal stability, $(\alpha-4)\approx 0$, however, the effective
thermal timescale becomes very long compared to the accretion time.
Equations (\ref{eq:pert1}) and (\ref{eq:pert2}) in this limit are
\begin{equation}\label{eq:pert3}
{\partial\over\partial t}\left({\delta T\over T}\right)\approx {2\over t_{\rm
therm}}{\delta y\over y}
\end{equation}
\begin{equation}\label{eq:pert4}
{\partial\over\partial t}\left({\delta y\over y}\right)\approx -{\alpha\over t_{\rm
accr}}{\delta T\over T}.
\end{equation}
These equations nicely summarize the physics of the
oscillations. Consider an upwards fluctuation in temperature $\delta
T>0$. At the point of marginal stability, the increase in heating rate
almost exactly cancels the $T^4$ dependence of the cooling rate. The
main effect is that the hotter temperature leads to faster burning of
the accreting fuel, and a decrease in thickness $\delta y<0$ on a
timescale $\sim t_{\rm accr}$ (eq.~[\ref{eq:pert4}]). But a thinner
layer cools faster since $F/y\propto 1/y^2$. Therefore, the
temperature fluctuation now begins to decrease, this time on the
faster timescale $\sim t_{\rm therm}$ (eq.~[\ref{eq:pert3}]). These
changes are out phase, driving an oscillation on the intermediate
timescale $\approx (t_{\rm therm}t_{\rm accr})^{1/2}$.

The behavior we describe here is shown by the canonical example of a
nonlinear oscillator, the van der Pol oscillator (e.g., Abarbanel,
Rabinovich, \& Sushchik 1993). This oscillator consists of an LC
circuit with an active element that can behave as a ``negative
resistor'', originally a vacuum tube. The governing equation is of the
form $\ddot x+k(x^2-1)\dot x+\omega^2 x=0$. Depending on the choice of
control parameter $k$, the behavior of this circuit is a limit cycle
(relaxation oscillations) with fast and slow timescales dominating at
different parts of the cycle ($k>1$), a strongly damped system which
evolves to a steady-state ($k<-1$), or oscillatory with growing or
damped oscillations ($|k|<1$). These three states are analogous to
bursting, stable burning, and oscillations near the stability boundary
in the one-zone model.

\subsection{Numerical integrations}

We have integrated equations (\ref{eq:heateqn}) and (\ref{eq:conteqn})
in time to determine the non-linear evolution of the one-zone
model. For the nuclear burning, we use the triple alpha reaction
($3\alpha\rightarrow ^{12}$C) rate as given by Fushiki \& Lamb (1987).
We allow for the presence of hydrogen in the accreted fuel, however,
by enhancing the energy release from the triple alpha reaction by a
factor $E_{\rm nuc}/E_{3\alpha}$, where $E_{3\alpha}=0.606$ MeV per
nucleon is the energy release from the triple alpha reaction, and
$E_{\rm nuc}$ is the energy release from burning the accreted mixture
of hydrogen and helium to iron group. We assume that $E_{\rm
nuc}=1.6+4.9X_0$ MeV per nucleon, where $X_0$ is the mass fraction of
hydrogen in the accreted layer. This expression for $E_{\rm nuc}$
includes an energy loss of 25\% from neutrino emission during
\textsl{$\alpha$p}- and \textsl{rp}-process burning (e.g., Fujimoto et
al.~1987), and gives $E_{\rm nuc}=5$ MeV per nucleon for $X_0=0.7$, in
good agreement with the energy release in the steady state burning
models of Schatz et al.~(1999).  We write the flux as $F=acT^4/3\kappa
y$ (Bildsten 1998), where the opacity $\kappa$ is calculated as
described by Schatz et al.~(1999). In addition, we include a flux
heating the layer from below of 0.15 MeV per nucleon, and a
contribution to the heating rate from hot CNO hydrogen burning in the
accumulating fuel layer.  Neither of these extra contributions to the
heat balance make a significant difference to our results.  Note that
the amount of hydrogen burned by hot CNO cycle prior to helium
ignition is very small at the rapid accretion rates considered here.

\begin{figure}
\plotone{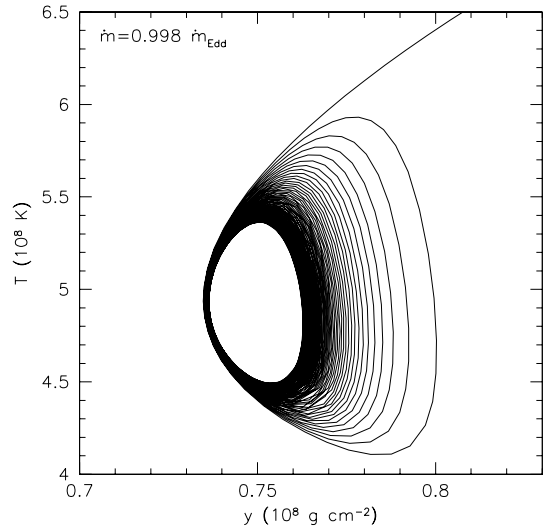}
\caption{The trajectory in the temperature-column depth plane for
$\dot m=0.998\dot m_{\rm Edd}$, as shown in Figure 2, but zooming in
on the oscillations around the steady burning
location.\label{Fig:tra1mag}}
\end{figure}

Figure~\ref{Fig:lc1} shows lightcurves from the one-zone integrations
at different accretion rates. To enable a direct comparison with the
multizone simulations discussed in \S 3, we set the local gravity to
be the Newtonian gravity for a 1.4 $M_\odot$, 10 km neutron star,
$g=1.9\times 10^{14}\ {\rm cm^2\ s^{-1}}$. Throughout this paper, we
define the local Eddington accretion rate to be $\dot m_{\rm
Edd}\equiv 8.8\times 10^4\ {\rm g\ cm^{-2}\ s^{-1}}$. By
coincidence\footnote{The physics which stabilizes the burning, the
decreasing temperature sensitivity of the triple alpha reaction with
increasing temperature, has nothing to do with the physics setting the
Eddington luminosity. By coincidence, the transition to stability
occurs close to the Eddington accretion rate (e.g., Bildsten 1998).},
the stability boundary for this one zone model is very close to the
Eddington accretion rate. In Figure~\ref{Fig:lc1}, we show lightcurves
at $\dot m=0.95$, $0.998$, and $1.05\ \dot m_{\rm Edd}$.
Figure~\ref{Fig:tra1} shows the corresponding tracks in the
temperature-column depth plane. We start the simulations with the
arbitrary conditions $T=2\times 10^8\ {\rm K}$, and $y=2\times 10^8\
{\rm g\ cm^{-2}}$. At accretion rates below the boundary, the system
evolves quickly into a limit cycle corresponding to Type I X-ray
bursts: slow accumulation of fuel followed by rapid burning. Close to
the stability boundary, the recurrence time is $\approx 30$
minutes. At accretion rates above the boundary the evolution is to a
steady state in which the fuel burns at $T\approx 5\times 10^8\ {\rm
K}$ and $y\approx 7\times 10^7\ {\rm g\ cm^{-2}}$. For accretion rates
very close to the transition, $\dot m\approx 1\ \dot m_{\rm Edd}$, we
find oscillations around the steady state conditions, with oscillation
period $\approx 4$ minutes and amplitude a few percent of the
Eddington flux. For the oscillatory case, Figure~\ref{Fig:tra1mag}
shows a more detailed view of the trajectory of the solution in the
temperature-column depth phase space.

\begin{figure*}
\epsscale{1.0}
\plotone{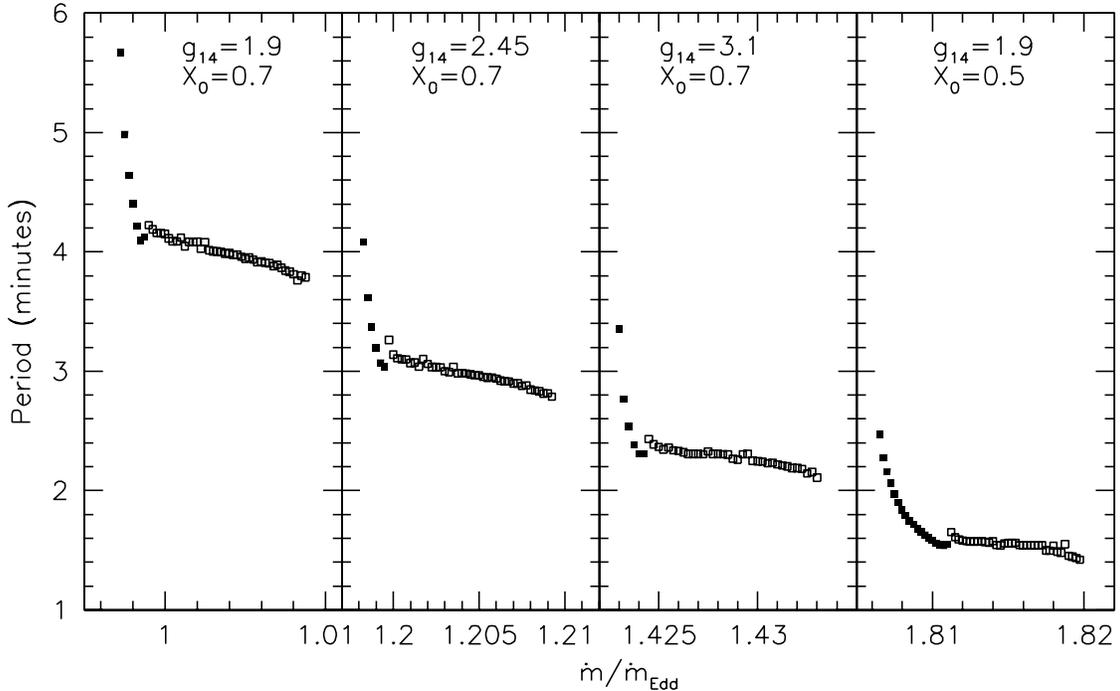}
\epsscale{1.0}
\caption{Oscillation period as a function of accretion rate in the
one-zone model, for different choices of surface gravity and accreted
hydrogen fraction. At each accretion rate, we integrate the model for
$10^6\ {\rm s}$, and plot the mean oscillation period, only including
data for $t>100$ minutes. The solid symbols indicate models for which
the oscillations grow and reach a steady amplitude; the open symbols
indicate models for which the oscillations are damped.\label{Fig:P1}}
\end{figure*}

Although the one-zone model is approximate, it is useful because it
allows us to investigate how the properties of the oscillations change
with parameters such as surface gravity and accreted composition. The
accreted composition could differ from system to system either due to
metallicity variations, or variations in the accreted hydrogen
fraction. Intermediate-mass binary evolution models
(e.g., Podsiadlowski et al.~2002) predict that the companion star in
many systems is hydrogen deficient, so that the hydrogen mass fraction
is reduced below the solar composition value of $X_0\approx 0.7$. The
importance of the hydrogen fraction is that the nuclear energy release
$E_{\rm nuc}$ changes significantly with only small changes in
$X_0$. In contrast, we expect that metallicity will not have a large
effect on the transition accretion rate or the oscillation period. This is because the
metallicity of the accreted material mainly enters into the one-zone
model as hot CNO hydrogen burning in the accumulating layer, but
the flux from hot CNO burning, $\epsilon_{\rm CNO}y\approx
5\times 10^{21}\ {\rm erg\ cm^{-2}\ s^{-1}}\ y_8(Z_{\rm CNO}/0.01)$,
is much smaller than the steady burning flux, $\dot mE_{\rm
nuc}\approx 5\times 10^{23}\ {\rm erg\ cm^{-2}\ s^{-1}}\ (\dot m/\dot
m_{\rm Edd})$, where $Z_{\rm CNO}$ is the mass fraction of CNO
elements.

Figure~\ref{Fig:P1} shows the dependence of the oscillation period on
accretion rate for different choices of gravity and accreted hydrogen
fraction. We show results for $g_{14}=1.9$, corresponding to the
Newtonian gravity of a 1.4 $M_\odot$, 10 km neutron star (or the
general relativistic gravity for a 1.4 $M_\odot$, 12.3 km neutron
star), $g_{14}=2.45$, the gravity of a 1.4 $M_\odot$, 10 km neutron
star taking general relativistic corrections into account, and a
stronger gravity $g_{14}=3.1$ which corresponds to the general
relativistic surface gravity for a $2\ M_\odot$, 11 km neutron
star. We also show a model with a hydrogen fraction below solar,
$X_0=0.5$. Only the accretion rate range at which oscillations are
observed is shown. At each accretion rate, we integrate the one-zone
model for $10^6$ seconds, and plot the mean oscillation period after
discarding the first 100 minutes of data.

For each choice of $g_{14}$ and $X_0$, the pattern is similar. The
overall range of accretion rates for which oscillations are seen is
very narrow, $\Delta \dot m/\dot m\approx 1$\%. For most of this range
the oscillations are decaying. As the stability boundary is approached
from below, we first see oscillations which reach a steady amplitude
(indicated by solid squares in Fig.~\ref{Fig:P1}) whose frequency
drops rapidly with increasing $\dot m$. At larger accretion rates
(open squares in Fig.~\ref{Fig:P1}), the oscillations decay with time,
on timescales $<1$ day, and have a frequency that is less sensitive to
$\dot m$. The transition from growing to decaying oscillations is very
rapid. In each case, the model indicated by the last closed square
shows stable oscillations for $10^6$ seconds, whereas with only a
small increment in accretion rate the next model (first open square)
has oscillations which decay on a timescale of $\sim 10^5$
seconds. Most interesting is that the oscillation period is sensitive
to $g_{14}$ and $X_0$. Increasing gravity or decreasing $X_0$ moves
the transition from unstable to stable burning to higher accretion
rates, where the oscillation period is shorter. As $X_0$ decreases,
the accretion rate range over which the oscillations are growing
rather than decaying is larger.

\section{Multizone calculations of burning near the stability boundary}

We now present detailed multi-zone models of nuclear burning at
accretion rates close to the transition from unstable to stable
burning. These models are extensions of the calculations presented by
Woosley et al.~(2004) using the implicit 1D hydrodynamic code
\textsc{Kepler} (Weaver et al.~1978). Woosley et al.~(2004) calculated
sequences of X-ray bursts at accretion rates $\dot M\approx 0.03$ and
$0.1\ \dot M_{\rm Edd}$.  In this paper we show the first results of
an extension of these calculations to higher accretion rates.

Following Woosley et al.~(2004), we take the gravitational mass of the
neutron star to be $1.4\ M_\odot$ and $R=10\ {\rm km}$, giving a
Newtonian gravity $g_{14}=1.9$. The effects of general relativity are
not included in the simulations itself, but because the burning layer
is very thin the effects of general relativity are small over the
extent of the simulated burning layer and a local Newtonian frame is a
good approximation.  General relativity may be accounted for using
appropriate redshift factors, as discussed in \S 4.4 of Woosley et
al.~(2004). The results we provide here do not include these redshift corrections;
the simulated conditions apply for different combinations of neutron
star radius and mass that give the same surface acceleration in the
local frame, using appropriate scaling of surface area, accretion
rate, and luminosity.

Our code includes an adaptive nuclear reaction network that
automatically adjusts to include or remove isotopes as needed to
follow the details of the nucleosynthesis, out of a reaction rate
library of about 5,000 nuclei (Rauscher et al.~2002).  The
calculations presented here use up to 1300 different isotopes.  The
reaction rate library includes recent measurements and estimates of
critical nuclear reaction rates. The effect of uncertainties in these
rates are discussed in detail in Woosley et al.~(2004).  The energy
generation from the reaction network is directly and consistently
coupled into the implicit hydrodynamic solver.  The numerical grid
adaptively refines and derefines the Lagrangian grid to resolve
gradients, but in the hydrogen-rich layer in effect is essentially at
constant mass resolution, i.e., linear in column depth.  Accretion is
modeled by periodically adding an extra zone at the surface of the
star (of column depth $\approx 1.6\times 10^6\ {\rm g\ cm^{-2}}$)
(Woosley et al.~2004). We follow both the compositional and thermal
profiles of the layer, including radiative and convective transport,
with a time-dependent mixing length treatment for convection,
semiconvection, and thermohaline convection.

\begin{figure}
\includegraphics[width=\columnwidth]{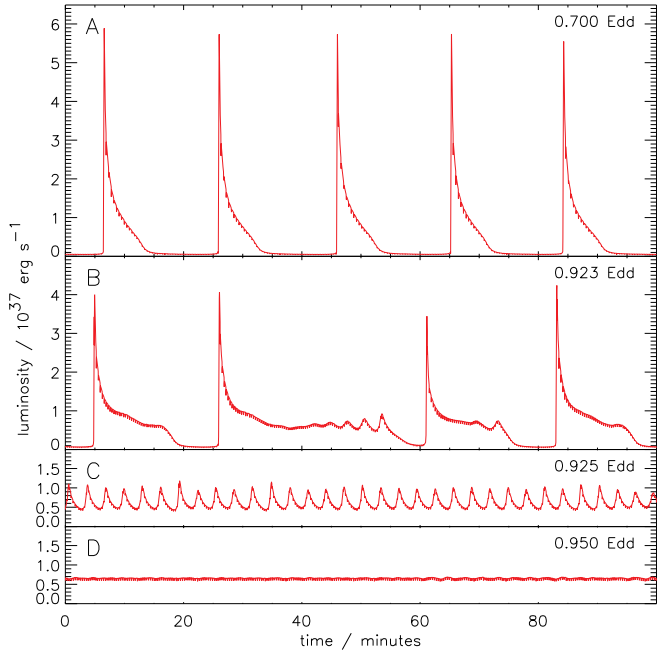}
\caption{Light curves for different accretion rates (the \textsl{upper
right corner} gives accretion rate in units of Eddington accretion
rate).  Panel~A shows regular bursting with stable recurrence times.
Panel~B shows weaker bursts with a partial oscillatory behavior in the
tail of the burst light curves.  Panel~C shows oscillatory rather
behavior and no bursts.  Panel~D shows very small oscillations,
essentially stable behavior.  The small dips seen in all lightcurves
on about 15 s time-scale are an artifact of our treatment of
accretion, in which an entire small outer zone is added
periodically. The beginning of this figure is 10,000 s after the
beginning of the simulation to allow the model to reach
quasi-stationary conditions.\label{Fig:lc}}
\end{figure}

Here we present results for an accreted material metallicity of 1/20
solar.  At an accretion rate $\dot M=0.1\ \dot M_{\rm Edd}$, Woosley et
al.~(2004) (their model zM) found a sequence of regular bursts with
recurrence times close to 3 hours. Increasing the accretion rate in
our new sequence of models, we find a transition from unstable to
stable burning at $\dot M=0.924\ \dot M_{\rm Edd}$.
Figure~\ref{Fig:lc} shows the behavior close to the transition
accretion rate. The lightcurves show a progression from regular
periodic bursting at $\dot M=0.7\ \dot M_{\rm Edd}$ (with recurrence
times close to $20$ minutes), to a combination of irregular bursts
intermixed with oscillations at $\dot M=0.923\ \dot M_{\rm Edd}$, to a
regular sequence of oscillations at $\dot M=0.925\ \dot M_{\rm Edd}$,
and finally to stable burning at $\dot M=0.95\ \dot M_{\rm Edd}$, with
the oscillation amplitude rapidly decreasing as the accretion rate is
increased.

\begin{figure}
\noindent
\includegraphics[bb=133 253 459 558,clip=true,width=\columnwidth]{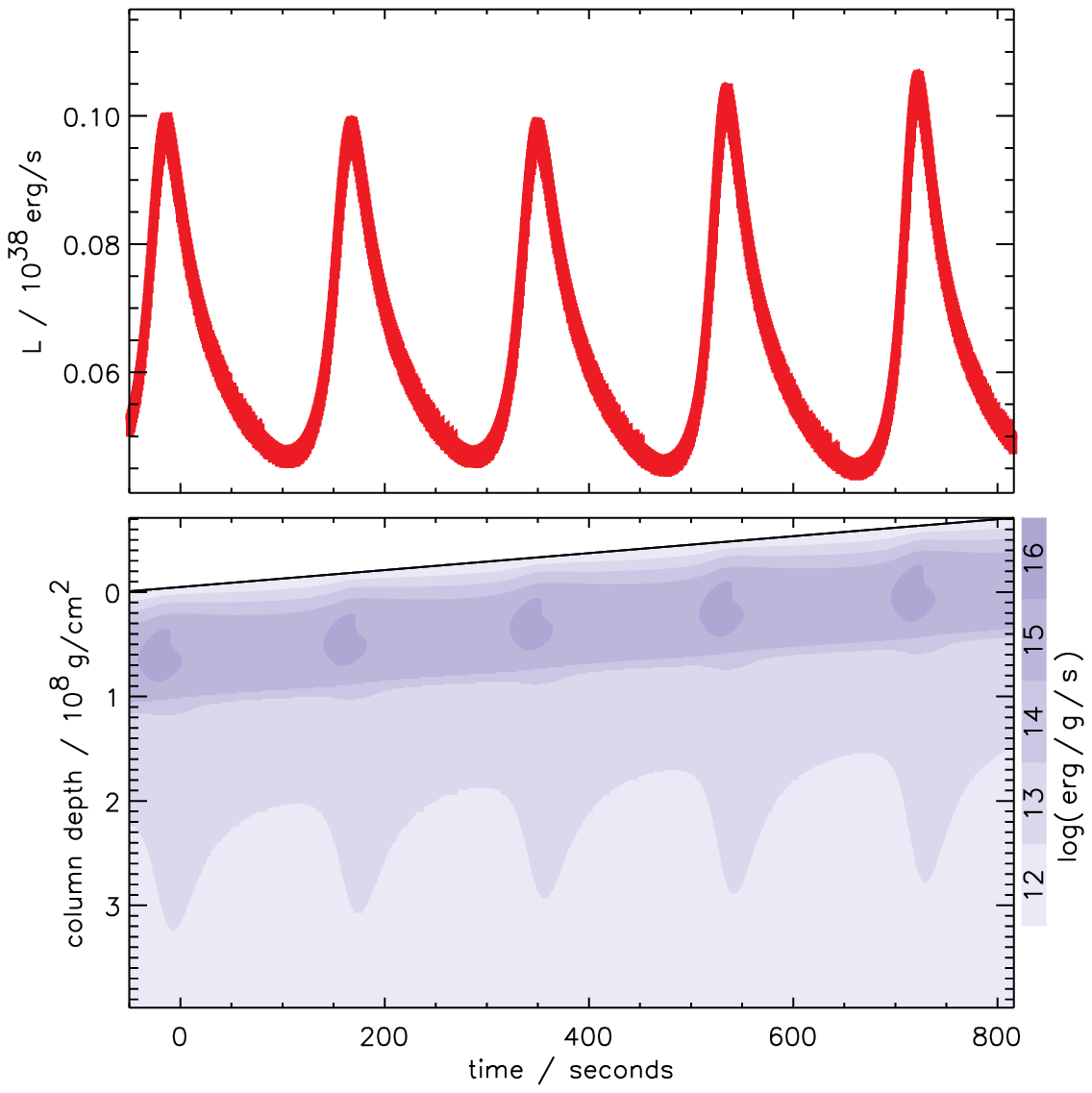}
\includegraphics[bb=135 325 459 486,clip=true,width=\columnwidth]{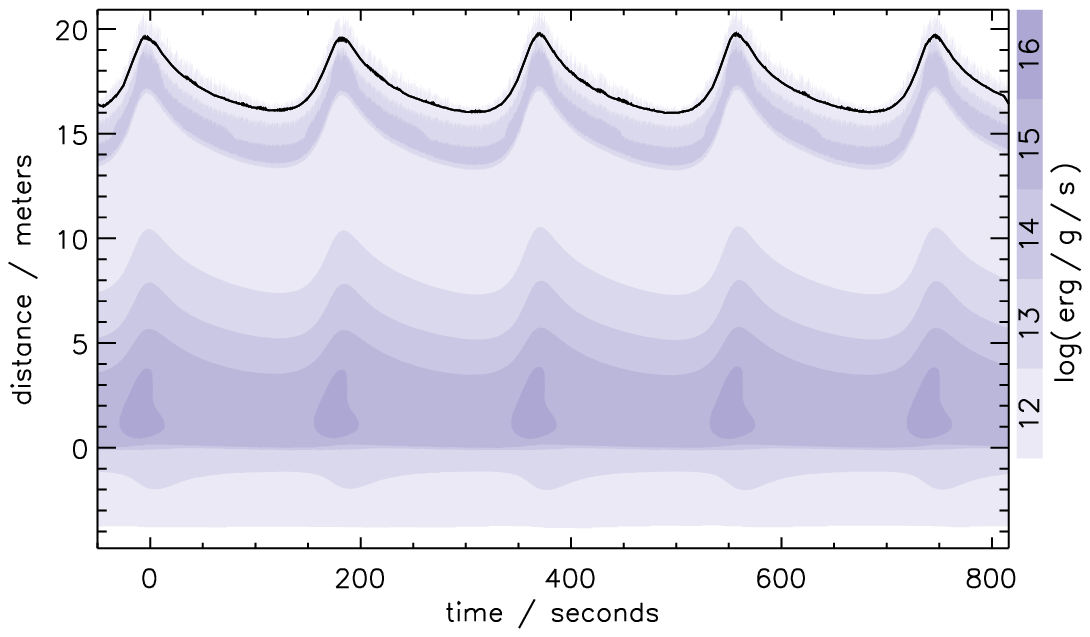}
\includegraphics[width=\columnwidth]{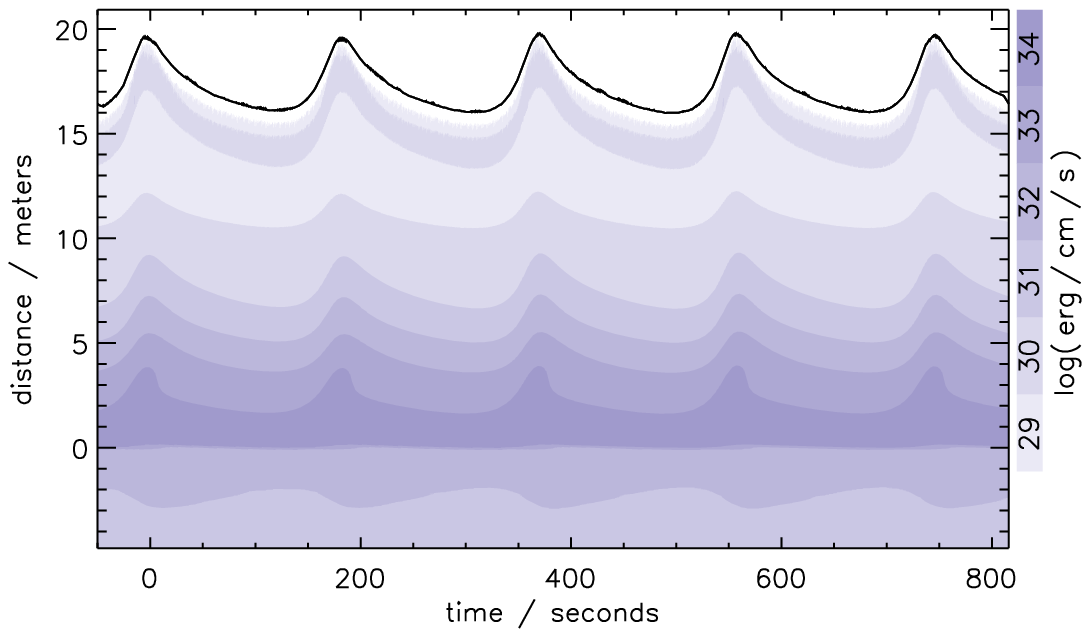}
\caption{Detailed light curve (\textsl{upper panel}) and specific
nuclear energy generation as a function of time and column depth
(\textsl{lower panel}).  In the lower thee panels, each darker shading
of blue corresponds to a value of energy generation one order of
magnitude higher; see scale on right hand side of the figures.  In the
second figure from the top we label each depth with a Lagrangian
column depth.  Following a given column depth to the right shows the
evolution of that fluid element in time.  The sloping black line
indicates the surface of the star (the slope gives the accretion
rate).  The lower two panels show the evolution s a function of radius
coordinate.  Zero is chosen to correspond to the location where
hydrogen is depleted.  The upper panel gives specific nuclear energy
generation rate (same as the panel above), the bottom panel gives
energy generation rate per unit depth.\label{Fig:kd}}
\end{figure}

The oscillation period at $\dot M=0.923\ \dot M_{\rm Edd}$ is $185\pm
5$ seconds. Figure~\ref{Fig:kd} shows a portion of the lightcurve at
this accretion rate. The oscillations have an asymmetric profile, with
the decay lasting twice as long as the rise. Revnitsev et al.~(2001)
noted marginal evidence that the peaks of the mHz QPOs were
asymmetric, with a steep rise and shallower decline. A more detailed
comparison of our models with observed lightcurves would be
interesting, but clearly there should be significant harmonic
components.  The peak-to-peak amplitude of the oscillation in
Figure~\ref{Fig:kd} is $\approx 5\times 10^{36}\ {\rm erg\ s^{-1}}$,
with minimum luminosity $\approx 5\times 10^{36}\ {\rm erg\ s^{-1}}$
and maximum luminosity $\approx 10^{37}\ {\rm erg\ s^{-1}}$. This is
in good agreement with the one-zone model (compare the middle panel of
Fig.~\ref{Fig:lc1}; for a 10 km star, a flux of $\approx 10^{24}\ {\rm
erg\ cm^{-2}\ s^{-1}}$ corresponds to a luminosity of $\approx
10^{37}\ {\rm erg\ s^{-1}}$).

\begin{figure}
\includegraphics[width=\columnwidth]{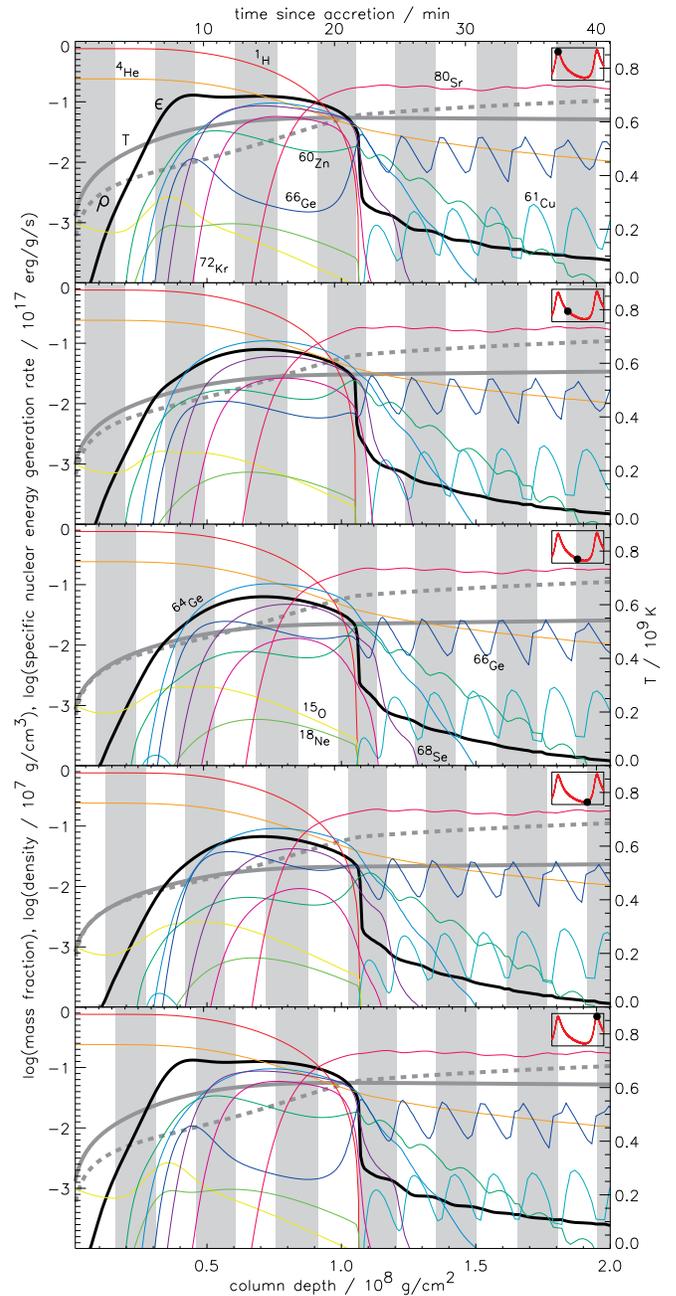}
\caption{Snapshots of structure (temperature: \textsl{thick gray
line}; density: \textsl{thick gray dashed line}; specific nuclear
energy generation: \textsl{black line}) and composition (select
isotopes, \textsl{colored lines}) during one oscillation cycle; each
panel is advanced in time by $P/4$ where $P$ is the oscillation
period; the bottom panel is advanced by one full cycle.  The bottom
axis for each figure gives column depth, the top axis the
corresponding time since accretion began.  This is the same model as
shown in Fig.~\ref{Fig:lc}, Panel~C.  The white and gray stripes
correspond to one cycle of oscillation each, with the interfaces
corresponding to the time of a maximum in the light curve at the time
of the accretion of that layer.  The small inserts at upper right
corners indicate the position in the light curve (\textsl{red}) cycle
of the snapshot (\textsl{black dot}; intentionally aligned with a
layer interface).  Note that the decreases of some of the radioactive
isotopes in the right-hand side of the figure is due to their
radioactive decay.  An animation of this figure can be found at
\texttt{http://xrayburst.org/qpo}.
\label{Fig:phase}}
\end{figure}

\begin{figure}
\includegraphics[angle=90,width=\columnwidth]{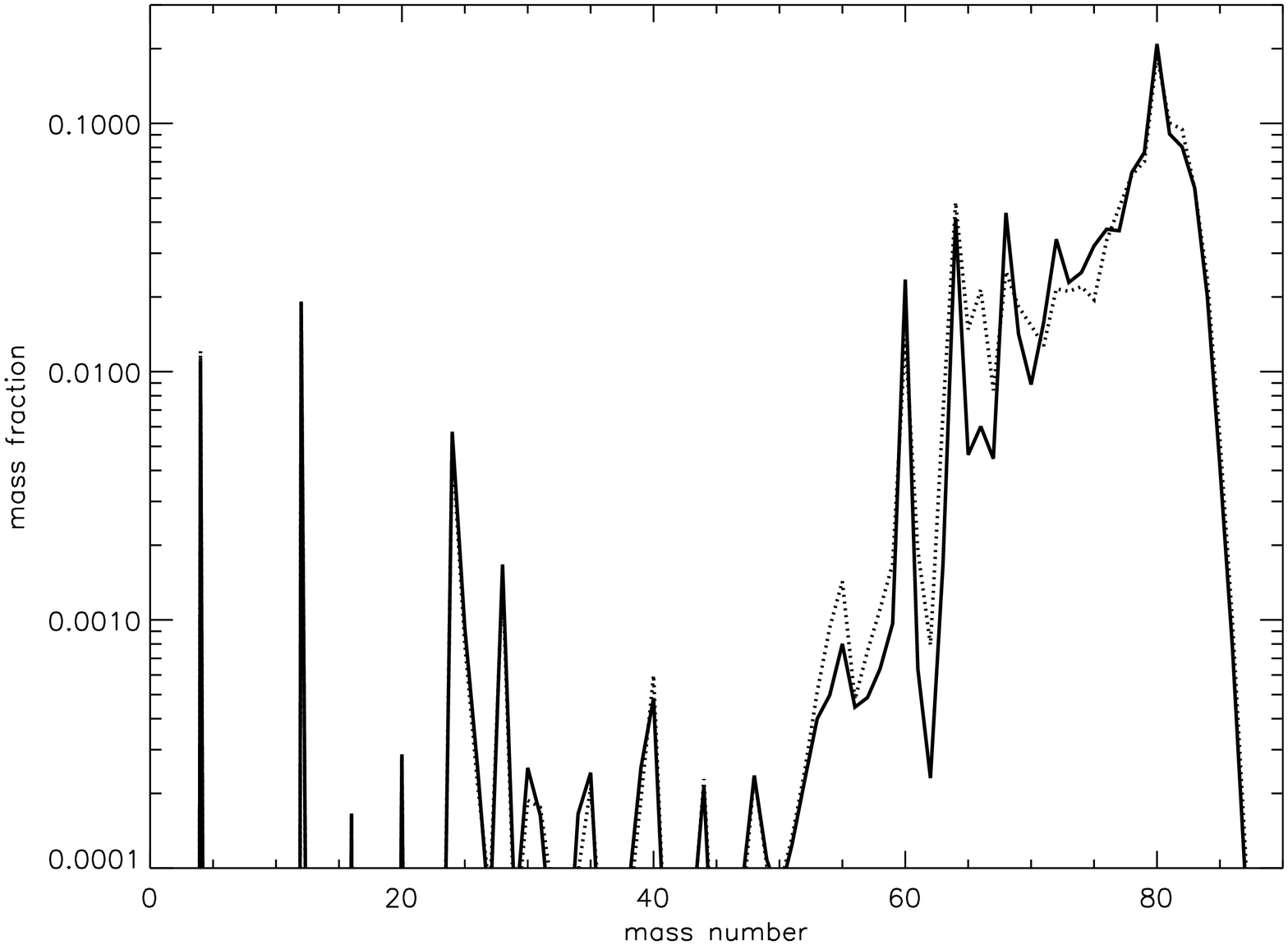}
\caption{Variation of the composition (summed up for each mass number
to be invariant against $\beta^+$ decays) in the ashes layer at a
column depth close to $2\times10^8\,$g$\,$cm$^{-2}$ (at the base of
the composition profile shown in
Fig.~\ref{Fig:phase}).\label{Fig:comp}}
\end{figure}

The nuclear burning in the oscillation mode is powered by
\textsl{$\alpha$p}- and \textsl{rp}-process burning (Wallace \&
Woosley 1981; Schatz et al.~1999), beginning with seed nuclei produced
by breakout reactions from the CNO cycle, and terminating at mass
number $\approx 80$ (the most abundant nucleus is
$^{80}$Sr). Figure~\ref{Fig:kd} shows the energy generation as a
function of column depth, radius, and time through several oscillation
cycles.  Figure~\ref{Fig:phase} shows the composition profile of the
layer at different phases of the oscillation cycle.  Beneath the
hydrogen burning layer, the composition of the ashes shows periodic
variations with a spacing in column depth of $\dot m P_{\rm osc}$,
where $P_{\rm osc}=185$ seconds is the oscillation period. Note,
however, that the hydrogen burning depth is relatively constant during
the oscillation cycle. The hydrogen burns at a depth $\approx 10^8\
{\rm g\ cm^{-2}}$ which is $\approx 7$ times larger than the amount of
mass accumulated in one oscillation cycle. The underlying ashes record
the variations in burning temperature and the resulting variation of
the rp-process ashes during the oscillation cycle.  This is
reminiscent of the growth of annual rings in a tree trunk.
Figure~\ref{Fig:comp} shows the distribution of nuclei in the ashes by
mass number, and the slight variation in the composition through the
cycle.

At the peak in the oscillation light curve, the increased temperature
in the burning region drives heat transport inwards, heating the
underlying material. This is similar to the substrate heating during
Type I X-ray bursts discussed by Woosley et al.\ (2004), and leads to
increase of burning of the $^4$He remaining in the ashes layer by the
triple-$\alpha$ process and by $\alpha$ captures.  In
Figure~\ref{Fig:kd} this can be seen as additional spikes in nuclear
energy generation below the main burning band of the hydrogen-rich
zone. Of particular interest is the amount of carbon remaining in the
ashes. Stable burning of accreted H/He has been suggested as the
source of carbon fuel for superbursts (in 't Zand et al.~2003; Schatz
et al.~2003). Figure~\ref{Fig:comp} shows that the amount of carbon at
$y\approx 2\times 10^8\ {\rm g\ cm^{-2}}$ is $\approx 2$\,\% by mass,
which is very close to the asymptotic value of slightly below $2$\,\% in
the deeper layers where all the helium has been burnt.

The layering of different compositions in the ashes does not persist
to great depths. The different layers have different values of the
number of electrons per baryon $\Ye$, which determines the specific
weight of fluid elements under the degenerate conditions in the ashes
layer. The variation in composition is stable to the Rayleigh-Taylor
instability because of the thermal buoyancy.  Secular doubly-diffusive
instabilities, however, cannot be suppressed.  We find that the
thermohaline or salt-finger instability grows in the layers which have
an outwardly-decreasing $Y_e$ profile. The subsequent mixing is
apparent in Figure~\ref{Fig:phase} as the ``erosion'' of the peaks in
the $^{68}$Ge profile and the valleys of the $^{61}$Cu profiles at
column depths above 1.5$\times10^{8}\,$g cm$^2$. The mixing eventually
leads to homogenization of the ashes layer at depths $>3\times 10^8\
{\rm g\ cm^{-2}}$. Our calculations followed that process to a point
where the typical column-depth scale of the homogenized regions was
several times that of the composition oscillations made by the
oscillations in the burning region.

Schatz et al.~(1999) calculated the nucleosynthesis in steady-state
burning models at $\dot m=1\ \dot m_{\rm Edd}$. They found that the
rp-process endpoint terminated at $A\approx 70$, and the carbon mass
fraction was $\approx 5$\%. We find slightly heavier rp-process ashes
($A\approx 80$), and a smaller carbon mass fraction ($\approx
2$\%). This is consistent with the general anti-correlation found by Schatz et al.~(2003) between
the mass of nuclei produced in the rp-process and the carbon yield. In our case, a longer rp-process gives
more time for helium to burn before the hydrogen runs out, leading to
less carbon production following hydrogen exhaustion. Our burning
temperature is approximately 10--20\% hotter than the models of Schatz
et al.~(1999) which might explain the more extensive rp-process. This
is perhaps because of a difference in radiative opacities: Schatz et
al.~(1999) use a fit to the results of Itoh et al.~(1991) for
free-free opacity, whereas the \textsc{Kepler} code uses the fit of
Iben (1975) to the radiative opacities of Cox \& Stewart (1970a,b). We
will investigate this difference further in future calculations.

\section{Discussion}

The fact that oscillatory burning is naturally expected at the
transition from unstable to stable nuclear burning was pointed out by
Paczynski (1981). We have investigated the properties of marginally
stable burning in this paper with a simplified one-zone model (\S2)
and with detailed multizone simulations (\S3) using the
\textsc{Kepler} code, extending the Type I X-ray burst calculations of
Woosley et al.~(2004) to higher accretion rates. The period,
amplitude, and shape of the oscillations agree well in both
models. Remarkably, the basic physics of the oscillations is the same
physics underlying a nonlinear relaxation oscillator such as the van
der Pol oscillator (e.g., Abarbanel et al.~1993). Usually, the
positive or negative damping term dominates, giving rise to the
familiar X-ray bursts at low accretion rates or stable burning at a
fixed temperature and density at high accretion rates.  Close to the
marginally stable point, however, the effective thermal timescale is
very long, allowing the underlying oscillation period of the system to
be seen. This period is close to the geometric mean of the thermal
time and accumulation time of the burning layer (eq.~\ref{eq:posc}).

\begin{figure}
\includegraphics[bb=73 210 556 672,width=\columnwidth]{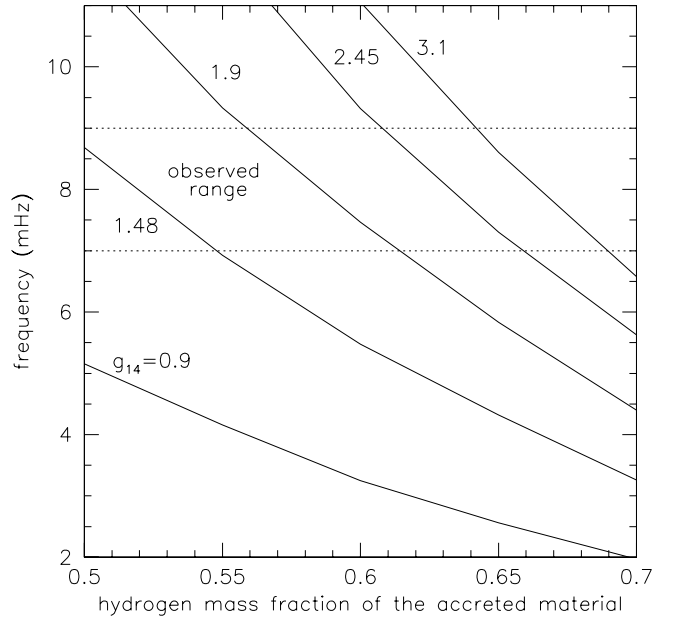}
\caption{Oscillation frequency as a function of hydrogen mass
fraction, $X_0$, for different surface gravities, corresponding to
different radii and masses of the underlying neutron star.  The data
is taken from one-zone models and rescaled to the oscillation
frequency found in the multi-zone model with $X_0=0.7$ and
$g_{14}=1.9$.  Proper general relativistic redshifts corrections and
radii have been applied.  The curve for $g_{14}=0.9$ corresponds to
$M=1.4\,M_\odot$ (gravitational mass) and $R=15.2\,$km, the curve for
$g_{14}=1.48$ corresponds to $M=1.4\,M_\odot$ and $R=11.9\,$km,
$g_{14}=1.9$ (used in the multi-zone calculation) corresponds to
$M=1.4\,M_\odot$ and $R=12.3\,$km, the curve for $g_{14}=2.45$ is for
$M=1.4\,M_\odot$ and $R=10\,$km, and the curve for $g_{14}=3.1$
corresponds to $M=2\,M_\odot$ and $R=11\,$km. The \textsl{dotted
lines} indicate the boundaries of the observed range of QPO
frequencies. \label{Fig:F(X)}}
\end{figure}

This behavior naturally reproduces two properties of the mHz QPOs
observed by Revnivtsev et al.~(2001): the observed periods of $\approx
2$ minutes, and the fact that mHz QPOs were observed in only a narrow
range of luminosities in 4U~1608-52, $(0.5$--$1.5)\times 10^{37}\ {\rm
erg\ s^{-1}}$. Identification of the mHz QPOs with marginally stable
nuclear burning would for the first time relate a feature of the
persistent X-ray emission to the neutron star surface in sources which
are not X-ray pulsars. Further, our one-zone models indicate that the
oscillation period is very sensitive to the surface gravity and
accreted hydrogen fraction (Fig.~\ref{Fig:P1}). One
of the difficulties in comparing X-ray burst properties with
theoretical models  is the uncertain relation
between X-ray luminosity and accretion rate (e.g., Cumming 2003). This uncertainty is
removed for marginally stable burning because it occurs at a specific
accretion rate. The dependencies on surface gravity and hydrogen
fraction need to be confirmed with multizone models, although unfortunately scanning the parameter space for the location of the transition is computationally very expensive.  Our one-zone model
results (Fig.~\ref{Fig:P1}) suggest that the observed periods of $\approx 2$ minutes require either $X_0<0.7$, as predicted by intermediate mass evolution
models (e.g., Podsiadlowski et al.~2002), or surface gravity
$g_{14}\approx 3$, for example corresponding to a $2\ M_\odot$
(gravitational mass) star with $R=11\ {\rm km}$. This can be seen in Figure~\ref{Fig:F(X)}, which shows the oscillation period as a function of $X_0$ and $g_{14}$. In this figure, we have rescaled the one-zone model results to match the multizone model for $X_0=0.7$ and $g_{14}=1.9$; in addition, we include the gravitational redshift factor\footnote{Gravitational redshift increases the oscillation periods shown in Figure~\ref{Fig:P1} by $\approx 30$\%. The numerical results in \S3, however, suggest that
the one-zone model overpredicts the oscillation period by a similar
factor. Therefore, the oscillation periods from the one-zone model without redshifting  are in fact approximately those we expect from multizone
models corrected for gravitational redshift.}. Future comparisons of theoretical models with
mHz QPO periods and lightcurves are potentially sensitive probes of
the surface gravity and accreted composition.

Two observed features of mHz QPOs remain to be explained.  The first
is the $Q$ value of the oscillation, which Revnivtsev et al.~2001
found to be $Q\equiv\nu/\Delta\nu\approx 3$--$4$. This may be related
to the range of accretion rates for which oscillatory nuclear burning
can be observed. Revnivtsev et al.~(2001) constrained the range of
luminosities at which mHz QPOs are present to be $(0.5$--$1.5)\times
10^{37}\ {\rm erg\ s^{-1}}$ in 4U~1608-52.  In contrast, the
theoretical range where oscillations are seen is much smaller, within
a range $\Delta \dot m/\dot m\approx 0.01$ around the transition
accretion rate. Moreover, for much of this range, the oscillations
decay in time. Further observations which constrain the range of
luminosities for which mHz QPOs can be observed would be valuable.

The second puzzle is that our theoretical models, in agreement with
previous estimates (Fujimoto, Hanawa, \& Miyaji 1981; Ayasli \& Joss
1982; Bildsten 1998), find that the transition to stable burning
occurs at an accretion rate close to the Eddington rate ($\dot
M=0.924\ \dot M_{\rm Edd}$ for the model presented in \S 3). In
contrast, the X-ray luminosity at which the mHz QPOs are observed in
4U~1608-52 is $\approx 0.5$--$1.5\times 10^{37}\ {\rm erg\ s^{-1}}$,
implying an accretion rate ten times lower, $\dot M\approx 0.1\ \dot
M_{\rm Edd}$. The same factor of ten appears when comparing the
theoretical and observed oscillation amplitudes. The observed
amplitudes of mHz QPOs correspond to flux variations of $\approx
1$--$2$\% (Revnivtsev et al.~2001). As noted by Revnivtsev et
al.~(2001), this is consistent with the ratio of nuclear energy
release from burning the accreted material to gravitational energy
release from accretion. We find fractional amplitudes of the expected
few percent level in the theoretical models.  The absolute
peak-to-peak flux variation is a factor of ten larger than observed,
however, roughly $5\times 10^{23}\ {\rm erg\ cm^{-2}\ s^{-1}}$ or
$5\times 10^{36}\ {\rm erg\ s^{-1}}$. This is more than $30$\% of the
observed persistent luminosity.  Note that the oscillation amplitude
decreases rapidly with increasing accretion rate above the transition
accretion rate, so there does exist a small range of accretion rates
where the amplitude does match the observed value.

A simple explanation for both of these discrepancies is that the {\em
local} accretion rate onto the star is close to the Eddington rate,
even though the {\em global} accretion rate is much lower. Because the
burning layer is very thin, the properties of the burning depend only
on the local accretion rate, which may vary across the surface of the
star. In particular, if the accreted material spread over only
$\approx 10$\% of the surface, the local accretion rate would be ten
times larger, comparable to the Eddington rate, and the emitted
luminosity would be ten times lower than if the burning covered the
entire stellar surface. This would allow marginally stable burning to
occur at a global rate of $\approx 0.1$ Eddington, while giving flux
variations of a few percent as observed.

The physics that might cause confinement of the fuel onto a small
fraction of the surface is not obvious. The pressure at the base of
the burning layer is $P=gy=10^{22}\ {\rm erg\ cm^{-3}}\ g_{14}y_8$,
implying that magnetic fields of strength approaching $\approx
\sqrt{8\pi P}\approx 3\times 10^{11}\ {\rm G}$ would be required to
confine the fuel. This is much larger than the $\approx 10^8$--$10^9\
{\rm G}$ fields assumed for the neutron stars in LMXBs (believed to be
the progenitors of the millisecond radio pulsars; Bhattacharya et
al.~1995), although small scale fields of these strengths might exist
on the surface. The need to transport angular momentum could also
potentially delay spreading of material accreted from a disk onto the
equator of the star. Inogamov \& Sunyaev (1999) studied this problem
with a one-zone model of the spreading layer and a basic prescription
for angular momentum transport. They found column depths $<10^4\ {\rm
g\ cm^{-2}}$ in the spreading layer, much smaller than the burning
depth.

The possibility that the covering fraction changes with accretion rate
was suggested previously by Bildsten (2000), but in the opposite
sense, with covering fraction increasing with $\dot m$. The motivation
was to explain a puzzling change in burst behavior that is observed to
occur at a luminosity $\sim 10^{37}\ {\rm erg\ s^{-1}}$. EXOSAT
observations of several Atoll sources showed that as X-ray luminosity
increased, burst properties changed from regular, frequent bursts
($t_{\rm recur}\approx$ hours) with energetics consistent with burning
all of the accreted fuel in bursts, to irregular, infrequent ($t_{\rm
recur}> 1$ day) bursts whose energetics indicate that only a small
fraction of the fuel burns in bursts (van Paradijs et al.~1988). RXTE
and BeppoSAX observations confirmed this result, with particularly
good coverage for the transient source KS~1731-260 (Muno et al.~2000;
Cornelisse et al.~2003). Cornelisse et al.~(2003) found that
observations of nine bursters with BeppoSAX were consistent with this
pattern of bursting, with a universal transition luminosity
$L_X\approx 2\times 10^{37}\ {\rm erg\ s^{-1}}$.

This change in bursting behavior is not predicted by the standard
theory, in which regular bursting should continue up to the stability
boundary at $\dot m\approx\dot m_{\rm Edd}$. Several theoretical
explanations for the discrepancy were put forward, including mixing by
Rayleigh-Taylor (Wallace \& Woosley 1984) or shear instabilities
(Fujimoto et al.~1987) which might allow more rapid burning of
hydrogen, a new mode of burning involving slowly propagating fires at
$\dot m\gtrsim 0.1\ \dot m_{\rm Edd}$ (Bildsten 1993), or that the
covering fraction of accreted material increases at higher accretion
rates, lowering the accretion rate per unit area and lengthening the
time between bursts, giving hydrogen time to burn stably (Bildsten
2000). We also mention here that Narayan \& Heyl (2003) calculated
linear eigenmodes of truncated steady-state burning models, and found
stability for accretion rates $\dot M>0.25\ \dot M_{\rm Edd}$, more
consistent with observations. The lower accretion rate is likely the
cause of the small oscillation frequencies that they found for
marginally stable burning (period $\sim 20$ minutes).  We find,
however, that bursting continues unabated up to $\dot M\approx \dot
M_{\rm Edd}$.  Further work comparing linear stability analysis with
numerical calculations seems to be required.

The observations of mHz QPOs at a luminosity close to $\approx
10^{37}\ {\rm erg\ s^{-1}}$ and their interpretation as marginally
stable oscillatory burning at a local accretion rate $\dot m\approx
\dot m_{\rm Edd}$ provide new input for these ideas. As we have
discussed, if changes in the covering fraction are responsible, this
suggests that the covering fraction {\em decreases} rather than
increases at this luminosity. The observed Type I bursts at
$L_X>10^{37}\ {\rm erg\ s^{-1}}$ could be accommodated if there was a
slow ``leak'' of fuel away from the stably burning region. This fuel
would deplete hydrogen as it accumulated, giving occasional short
helium-rich bursts. Other mechanisms, such as stable burning driven by
mixing of fuel by shear instabilities might also lead to oscillatory
burning. More theoretical work is needed. One clue is that the
ultracompact source 4U~1820-30 which most likely accretes pure helium
(Bildsten 1995; Cumming 2003) shows a similar transition at a similar
luminosity, implying that the nature of the transition does not depend
on accreted composition.

This question is also likely to be relevant for superbursts and Type I
burst oscillations. Superbursts are long duration, rare, and extremely
energetic Type I X-ray bursts (up to 1000 times the duration and
energy and less frequent than normal bursts) believed to be due to
unstable carbon ignition (Cumming \& Bildsten 2001; Strohmayer \&
Brown 2002). Superbursts are only observed at luminosities above
$\approx 10^{37}\ {\rm erg\ s^{-1}}$, and from sources for which burst
energetics indicate that bursts burn only a small fraction of the
accreted fuel (in 't Zand et al.~2003). This fits nicely with the
theoretical result that stable burning is much more efficient than
unstable burning at producing the carbon fuel (Schatz et al. 2003).
How to achieve stable burning theoretically at $\dot m<\dot m_{\rm
Edd}$, however, has been an open question. Type I burst oscillations
are high frequency oscillations during Type I X-ray bursts believed to
be due to burning asymmetries on the surface. Burst oscillations are
preferentially seen at higher accretion rates, in the banana branch of
the color-color diagram (e.g., Muno et al.~2000). Incomplete covering
of the surface might help to explain the origin of burst oscillations,
facilitating inhomogeneous burning when Type I X-ray bursts are able
to occur.

Finally, we note that Atoll sources undergo a transition from the
island state to banana branch in the color-color diagram at a
luminosity of $L_X\approx 10^{37}\ {\rm erg\ s^{-1}}$. It is
well-known for these sources that X-ray luminosity does not track
accretion rate on short timescales (van der Klis 2001).  One
explanation for the transition from island state to banana branch in
Atolls is that a hot quasi-spherical accretion flow at low rates is
replaced by a thin disk at high rates (e.g., Gierlinski \& Done 2002).
If this picture is correct, it could be that the change in accretion
geometry affects the distribution of fuel on the neutron star
surface. An interesting observational question is whether the
appearance of the mHz QPOs is linked to a particular luminosity range,
or a particular part of the color-color diagram, for example the
island to banana transition.

\acknowledgements We thank Michiel van der Klis and Dong Lai for useful
discussions.  This research is supported by the DOE Program for
Scientific Discovery through Advanced Computing (SciDAC;
DE-FC02-01ER41176).  AH is supported at LANL by DOE contract
W-7405-ENG-36 to the Los Alamos National Laboratory.  AC acknowledges
support from McGill University startup funds, an NSERC Discovery
Grant, Le Fonds Qu\'eb\'ecois de la Recherche sur la Nature et les
Technologies, and the Canadian Institute for Advanced Research.  SEW
has been supported by the NSF (AST 02-06111) and NASA (NAG5-12036).

\end{document}